\documentclass[12pt]{article}

\usepackage{amsmath}
\usepackage{amsfonts}
\usepackage{amssymb}
\usepackage{multicol}
\usepackage{color}
\usepackage{graphicx}
\usepackage{pstcol}
\usepackage{pst-coil}
\usepackage{colortbl}
\usepackage[all]{xy}

\newpsobject{showgrid}{psgrid}{subgriddiv=1,griddots=10,gridlabels=6pt}

\definecolor{dark-green}{rgb}{0,0.7,0}
\definecolor{dark-blue}{rgb}{0,0.2,0.5}
\definecolor{med-blue}{rgb}{0,0.7,1}
\definecolor{mblue}{rgb}{0,0.2,1}
\definecolor{cnc}{rgb}{0.8,0,0}
\definecolor{light-red}{rgb}{1,0.8,0.8}
\definecolor{dark-yellow}{rgb}{1,0.8,0}
\definecolor{light-blue}{rgb}{0.8,0.9,1}
\definecolor{verylight-blue}{rgb}{0.93,0.95,1}
\definecolor{light-yellow}{rgb}{1,0.9,0.8}
\definecolor{grey}{gray}{0.88}

\def\a{\alpha}

\begin{document}

\title{Linear media in classical electrodynamics and the Post
  constraint}

\author{Friedrich W.\ Hehl$^{1,2}$ and 
Yuri N.\ Obukhov$^{1,3}$
\\ \\ $^1$ Institut f\"ur Theoretische Physik, Universit\"at zu
  K\"oln\\ 50923 K\"oln, Germany   \\
  $^2$ Department of Physics and Astronomy, University of
  Missouri-Columbia\\ Columbia, MO 65211, USA\\
$^3$ Department of Theoretical Physics,
  Moscow State University\\ 117234 Moscow, Russia}

\date{3 November 2004, {\it file PostCon9.tex}}

\maketitle

\begin{abstract}
  The Maxwell equations are formulated in a generally covariant and
  metric-free way in 1+3 and subsequently in 4 dimensions. For this
  purpose, we use the excitations $\cal D$, $\cal H$ and the field
  strengths $E,B$. A local and linear constitutive law between
  excitations and field strengths is assumed, with a constitutive
  tensor $\chi^{ijkl}=-\chi^{jikl}=-\chi^{ijlk}$ of 36 components. The
  properties of this tensor are discussed. In particular, we address
  the validity of the Post constraint, a subject that is very much
  under discussion. In this connection, the Tellegen gyrator, the axion
  field, and the ``perfect electromagnetic conductor'' of Lindell \&
  Sihvola are compared with each other.
\end{abstract}

\noindent Keywords: Electrodynamics, linear medium, constitutive tensor, 
Post constraint, axion field, Tellegen gyrator, skewon field, classical
field theory \medskip

\noindent PACS: 03.50.De, 46.05.+b, 14.80.Mz



\section{Introduction}

Electrical engineers as well as physicists are familiar with the
Maxwell equations
\begin{eqnarray}
  \underline{d}\,{\cal D} &=& \rho\,,\qquad \underline{d}\,{\cal H} -
  {\dot{\cal D}} = j\,,\label{Minhom} \\ \underline{d}\,B &=&
  0\,,\qquad \underline{d}\,E\, + {\dot B} = 0\,,
\label{Mhom}
\end{eqnarray}
here expressed \cite{Bamberg,Frankel,Birkbook,Ismo,Russer} in exterior
differential forms in terms of the electric and magnetic excitations
${\cal D},{\cal H}$ and the electric and magnetic field strength
$E,B$, respectively.  With $\underline{d}$ we denote the 3-dimensional
exterior derivative and with a dot the time derivative. The sources of
the inhomogeneous equations are the electric charge density $\rho$ and
the electric current density $j$. These equations are expressed in a
generally covariant form, that is, they are valid in {\it arbitrary
  curvilinear coordinates}. Moreover, they are {\it metric-free\/} and
thus the concept of orthogonality is not used at all.

As shown in the literature, see \cite{Schouten,Post,Birkbook}, the
Maxwell equations can alternatively be written in terms of components
as
\begin{eqnarray}\label{Max3D}
  && \partial_a{\cal D}^a=\rho\,,\qquad \epsilon^{abc}\,\partial_b{\cal
    H}_c-\dot{{\cal D}}^a=j^a\,,\\ && \partial_a{B}^a=0\,,\qquad
  \epsilon^{abc}\,\partial_b{E}_c\,+\dot{{B}}^a=0\,,
\end{eqnarray}
with $a,b,\dots=1,2,3$ and $\epsilon^{abc}=\pm1,0$ as the totally
antisymmetric Levi-Civita symbol. Here ${\cal D}^a,B^a,j^a$ are vector
densities, $E_a,{\cal H}_a$ covectors, and $\rho$ is as scalar
density. Also in this formulation the general covariance and the
condition of being free of the metric are manifest.  Note, since no
metric is used, we must not raise and lower indices with the help of
the metric at this stage.

The Maxwell equations need to be supplemented by constitutive
relations. And this for the description (i) of empty spacetime (the
vacuum) and (ii) of material media.  We will discuss here the simplest
case, namely local and linear media.  Local in space as well as in
time, that is, the excitations at every point in space and time depend
only on the field strength at the very same point.

A general local and linear constitutive relation reads (in the
conventions of \cite{Birkbook}),
\begin{eqnarray}\label{const1}
  {\cal H}_a&=&-{\frak C}^{b}{}_a\,E_b+{\frak B}_{ba}\,B^b\,,\\ {\cal
    D}^a&=& -{\frak A}^{ba}\,E_b+{\frak
    D}_{b}{}^a\,B^b\,.\label{const2}
\end{eqnarray}
The generalized permittivity matrix ${\frak A}^{ba}$ and the
generalized {\it im\/}permeability matrix (the reciprocal of the
permeability) ${\frak B}_{ba}$ are well known. Less known, but well
established in the electrodynamics of continuous media, see
\cite{O'Dell}, are the magneto-electric matrices ${\frak C}^{b}{}_a$
and ${\frak D}_{b}{}^a$, which describe magnetic-electric cross
effects (like the Faraday effect or optical activity).  Altogether,
these matrices encompass $4\times 9=36$ independent components. Since
$ {\cal H}_a,{\cal D}^a$ and $E_b,B^b$ are real, the matrices
$\frak{A}^{ba}$ etc.\ are also real. The subject of our article will
be the discussion of certain possible algebraic constraints between
the components of the constitutive matrices ${\frak A}^{ba},{\frak
  B}_{ba},{\frak C}^{b}{}_a,{\frak D}_{b}{}^a$.  The most notorious
one is the so-called Post constraint, which requires the sum of the
traces of the matrices ${\frak C}^{b}{}_a$ and ${\frak D}_{b}{}^a$ to
vanish, i.e., ${\frak C}^{a}{}_a+{\frak D}_{a}{}^a=0 $.  There is an
ongoing dispute mainly in the electrical engineering community, see
Lakhtakia \& Weiglhofer
\cite{Akhlesh2004a,Akhlesh2004b,LakhWeig1994,LakhWeig1995,LakhWeig1996,Weig1994,WeigLakh1998}
and Lindell, Sihvola, Tretyakov, et al.\ 
\cite{LindSihv2004a,LindSihv2004b,Lindell1994,LindellWallen2004a,LindellWallen2004b,RaabSihvola1997,SihvolaLindell1995,Tretyakov1998}
whether this constraint is valid or not. We will address this
question.

The constitutive matrices are functions of space $x^a$ and time $t$ in
general.  In other words, they are {\it fields\/} ${\frak
  A}^{ba}={\frak A}^{ba}(x^a,t)$, etc. For applications to the case
(i), to empty spacetime, they can be constants if a flat
Minkowski spacetime and Cartesian coordinates are prescribed.
However, if a Riemannian spacetime of general relativity is assumed or
just a 4-dimensional differential manifold, then clearly the
constitutive matrices are fields. Similarly, in case (ii), for
material media, a homogeneous medium requires constant constitutive
matrices, as soon as inhomogeneous media are allowed, the constitutive
matrices are fields again.  We may call them generalized
permittivity/impermeability fields.

Before we turn to a 4-dimensional discussion of the Maxwell equations,
we want to comment on the constitutive relations
(\ref{const1}),(\ref{const2}). In the next section we will show that for
reasons of 4-dimensional covariance (\ref{const1}),(\ref{const2})
represent the appropriate constitutive relations. However, in the
engineering as well as in the physics literature, we often find $
{\cal D}^a$ and $B^a$ expressed in terms of $E_b$ and $H_b$. If we
exclude singular cases, i.e.,
\begin{equation}\label{neu5}
{\rm det}\,{\frak B}\ne 0\,,
\end{equation}
then (\ref{const1}) can be multiplied with ${\frak B}^{-1}$, solved
with respect to $B^a$, and substituted into (\ref{const2}). In this
way, we find, suppressing indices,
\begin{eqnarray}\label{neu6}
  {\cal D}\!&=\!& \left(-{\frak A} + {\frak D}{\frak B}^{-1}{\frak C}
  \right)E+\left({\frak D}{\frak B}^{-1} \right){\cal H} \,,\\ 
  {B}\!&=\!&\hspace{43pt} \left({\frak B}^{-1}{\frak C}
  \right)E
+\hspace{9pt}\left({\frak B}^{-1} \right){\cal
    H}\,.\label{neu7}
\end{eqnarray}
Thus, the the $EB$-formalism in (\ref{const1}),(\ref{const2}) is
transformed to the $E{\cal H}$-formalism of (\ref{neu6}),(\ref{neu7}).
Both formalisms are equivalent, provided a singular case can be
excluded. The natural linear constitute relations are
represented by (\ref{const1}),(\ref{const2}), as was shown by Post
\cite{Post} and O'Dell \cite{O'Dell}, e.g., and as we will discuss in
the next section. Hence we will concentrate on
(\ref{const1}),(\ref{const2}) and will consider (\ref{neu6}),(\ref{neu7})
only as secondary.

\section{Electrodynamics in 4-dimensional spacetime}

The true structure of classical electrodynamics can be more clearly
recognized if we present it in a 4-dimensional way. If we define
excitation, electric current, and field strength according to
\begin{eqnarray}
  H &=& {\cal D}-\,{\cal H}\wedge dt \,,\qquad J =\rho
  -\,j\wedge dt\, \,,\label{inhMax'}\\  F
 & =&B+\,E\wedge dt \,,\label{homMax'}
\end{eqnarray}
then the 4-dimensional Maxwell equations (\ref{Minhom}) and
(\ref{Mhom}) read
\begin{eqnarray}
  d\,H&=&J\,,\label{inhMax}\\ d\,F&=&\,0\,.\label{homMax}
\end{eqnarray}
This shows explicitly that they are invariant under 4-dimensional
coordinate/frame transformations and independent of the metric of
spacetime, see \cite{Birkbook}. If we require metric independence, we
get a well-prescribed way of formulating the Maxwell equations. Since
the metric of spacetime is the gravitational potential in general
relativity, it is gratifying to know that there is a gravity-free way
of formulating the Maxwell equations.  Accordingly, the Maxwell
equations (\ref{inhMax}),(\ref{homMax}) are valid in a flat Minkowski
spacetime (in any coordinates), in a curved Riemannian spacetime, and
even in the Riemann-Cartan spacetime of the Poincar\'e gauge theory of
gravitation. Needless to say that this beautiful formalism has great
practical value for applying electrodynamics in accelerated reference
frames, for instance.

The excitation $H$ is a {\it twisted\/} 2-form, the field strength $F$
an {\it un\/}twisted 2-form, see the discussion of Post in
\cite{Post95}. We decompose them in their components according to
$H=H_{ij}\,dx^i\wedge dx^j/2$ and $F=F_{ij}\,dx^i\wedge dx^j/2$. Here
$x^i$ are spacetime coordinates with $i,j,\dots =0,1,2,3$.
Analogously, the current $J$ as a twisted 3-form decomposes as
$J=J_{ijk}\,dx^i\wedge dx^j\wedge dx^k/6$.  Thus, in the formalism of
tensor analysis, (\ref{inhMax}) and (\ref{homMax}) can alternatively
be written as
\begin{eqnarray}
  \partial_k\check{H}^{ik}&=&\check{J}^i \,,\label{inhMaxta}\\ 
  \partial_{[i}F_{kl]}&=&\,0\,,\label{homMaxta}
\end{eqnarray}
with $\check{H}^{ik}:=\epsilon^{ikl m}\,H_{l m}/2$ and
$\check{J}^i:=\epsilon^{ikl m}\,J_{kl m}/6$. Here $\epsilon^{ikl
  m}=\pm 1,0$ is the 4-dimensional contravariant Levi-Civita symbol.
The covariant Levi-Civita symbol, which we will use below, is denoted
by a circumflex: $\hat{{\epsilon}}_{ikl m}=\pm 1,0$.  Incidentally,
since no metric is available for the raising and lowering of indices
at this stage, we have to distinguish these two symbols.

In the constitutive law (\ref{const1}),(\ref{const2}), we assumed {\it
  locality\/} and {\it linearity.} In 4 dimensions this translates
into
\begin{eqnarray}\label{constitutive}
H=\kappa(F)\,,
\end{eqnarray}
where $\kappa$ is a local and linear operator. Accordingly,
\begin{equation}\label{constit} H_{ij} =
  \frac{1}{2}\,\kappa_{ij}{}^{kl}\,F_{kl}\qquad {\rm with}\qquad
  \kappa_{ij}{}^{kl}=-\, \kappa_{ji}{}^{kl} =-\,\kappa_{ij}{}^{lk}\,.
\end{equation}
Here $\kappa_{ij}{}^{kl}(x)$ is the twisted constitutive tensor of
type $\left[^{2}_{2}\right]$. It can characterize (i) either empty
spacetime or (ii) a material medium. It has at most 36 independent
components. In general, it is a field, that is, it can describe an
inhomogeneous spacetime or an inhomogeneous medium.

The relativistic invariance properties are crucial for deciding
between the $EB$- and $E{\cal H}$-formal\-isms.  {}From the point of
view of a 4-dimensional spacetime, $E$ and $B$ form the invariant
field strength $F$; likewise, ${\cal D}$ and ${\cal H}$ form the
invariant excitation $H$. Accordingly, the linear constitutive law
(\ref{const1}),(\ref{const2}) is a relativistically invariant
statement, as it is seen from (\ref{constitutive}). This law involves
the constitutive matrices $\frak A$, $\frak B$, $\frak C$, and $\frak
D$ that are combined into a covariant object -- the constitutive
tensor $\kappa_{ij}{}^{kl}$. In contrast, the pairs ${\cal D}$, $B$
and $E$, ${\cal H}$ do {\it not\/} form any 4-dimensional relativistic
objects, and, as a result, the constitutive relation in the form of
(\ref{neu6}), (\ref{neu7}) involves non-covariant objects with unclear
physical meaning.
Therefore, the $EB$-scheme of (\ref{const1}),(\ref{const2}) is, from a
relativistic point of view, to be preferred against the $E{\cal
  H}$-scheme of (\ref{neu6}),(\ref{neu7}), see O'Dell \cite{O'Dell},
Chap.2, Sec.1.2, pp.23--24. If $\det {\frak B}\ne 0$, both systems are
formally equivalent; still, (\ref{neu6}),(\ref{neu7}), as a historical
artifact, should be phased out from use.

One can decompose the constitutive tensor $\kappa_{ij}{}^{kl}$ into
its irreducible pieces under the linear group acting in the tangent
and the cotangent space, respectively. Contraction is the only tool
for such a decomposition.  Following Post \cite{Postmap}, we can
define the contracted twisted tensor of type $\left[^{1}_{1}\right]$
\begin{equation}
\kappa_i{}^k := \kappa_{il}{}^{kl}\,,
\end{equation}
with 16 independent components. 
The second contraction yields the twisted scalar function
\begin{equation}\label{trace}
\kappa := \kappa_k{}^k = \kappa_{kl}{}^{kl}\,
\end{equation}
(also called pseudo- or axial-scalar).  The traceless piece
\begin{equation}\label{tracefree}
\not\!\kappa_i{}^k := \kappa_i{}^k - {\frac 1 4}\,\kappa\,\delta_i^k
\end{equation}
has 15 independent components. These pieces (\ref{trace}) and
(\ref{tracefree}) can now be subtracted from the original constitutive
tensor. Then we find the following three irreducible pieces:
\begin{eqnarray}
  \kappa_{ij}{}^{kl} &=& {}^{(1)}\kappa_{ij}{}^{kl} +\nonumber
  {}^{(2)}\kappa_{ij}{}^{kl} + {}^{(3)}\kappa_{ij}{}^{kl} \\ &=&
  {}^{(1)}\kappa_{ij}{}^{kl} +
  2\!\not\!\kappa_{[i}{}^{[k}\,\delta_{j]}^{l]} + {\frac 1
    6}\,\kappa\,\delta_{[i}^k\delta_{j]}^l\,.\label{kap-dec}
\end{eqnarray}

By construction, ${}^{(1)}\kappa_{ij}{}^{kl}$ is totally traceless:
\begin{equation}
{}^{(1)}\kappa_{il}{}^{kl} = 0.\label{notrace}
\end{equation}
Thus, we split $\kappa_{ij}{}^{kl}$ according to $36 = 20 + 15 + 1$,
since the $\left[^{2}_{2}\right]$ tensor ${}^{(1)}\kappa_{ij}{}^{kl}$
is subject to the 16 constraints (\ref{notrace}) and hence carries $36
-16=20$ components.  We call ${}^{(1)}\kappa_{ij}{}^{kl}$ the {\it
  principal\/} (or the metric-dilaton) part of the constitutive
tensor.  Without such a term, the propagation of electromagnetic waves
is ruled out, see the discussion subsequent to
eq.(\ref{G4}). We further identify the two other irreducible
parts ${}^{(2)}\kappa_{ij}{}^{kl}$ and ${}^{(3)}\kappa_{ij}{}^{kl}$ as
{\it skewon\/} and {\it axion\/} fields, respectively. Conventionally,
the skewon and the axion fields are introduced with different factors
by
\begin{equation}
\!\not\!S_i{}^j := -\,{\frac 1 2}\!\not\!\kappa_i{}^j,\qquad
\alpha := {\frac 1 {12}}\,\kappa.\label{Salpha}
\end{equation}

If we substitute (\ref{kap-dec}) into (\ref{constit})$_1$ and use
(\ref{Salpha}), then the general linear constitutive law can be
written as
\begin{equation}\label{crypto2a}
  H_{ij}=\frac 12\,^{(1)}\kappa_{ij}{}^{kl}\,F_{kl}+2\, {\!\not
    \!S}_{[i}{}^kF_{j]k}+\a\,F_{ij}\,.
\end{equation} 
It has been shown recently \cite{L-H} that taking the linear spacetime
relation for granted, one can derive a Riemannian light cone provided
one forbids birefringence in vacuum.  However, if one allows for the
skewon field, then the light cone is dissolved and we find birefringence
and more general type of optical effects \cite{skewon}. 

Let us now turn to some examples for case (i), the vacuum.

\subsubsection*{Maxwell-Lorentz vacuum electrodynamics}

If we assume a Riemannian metric $g_{ij}$ with Lorentzian signature,
then vacuum electrodynamics is recovered by
\begin{equation}\label{MaxLor}
  {}^{(1)}\kappa_{ij}{}^{kl} =
  \lambda_0\,\sqrt{-g}\,\hat{\epsilon}_{ijmn}\,g^{mk} g^{nl}\,,\quad
  \!\not\!S_i{}^j=0\,,\quad \alpha=0\,.
\end{equation}
Here $\lambda_0 = \sqrt{\varepsilon_0/\mu_0}$ is the vacuum admittance
and $g:=\det g_{kl}$.  The principal part $^{(1)}\kappa_{ij}{}^{kl}$
corresponds to the Hodge star operator $\star$. The Maxwell equations
can then be written as\footnote{The inhomogeneous equation in indices
  reads $\lambda_0\,\partial_j\left(\sqrt{-g}g^{ik}g^{jl}F_{kl}
  \right)=\check{J}^i$.}
\begin{equation}\label{Max*}
\lambda_0\,d\,^\star F=J\,,\quad dF=0\,.
\end{equation}

\subsubsection*{Axion electrodynamics}
If we admit additionally the axion field $\alpha\ne 0$, then
$H=\lambda_0\,^\star F+\alpha F$. This framework is called { axion\/}
(Maxwell-Lorentz) { electrodynamics,} see Ni \cite{Ni73,Ni77,Ni84} and
Wilczek \cite{Wilczek87}, e.g.:
\begin{equation}\label{MaxAx}
\lambda_0\,d\,^\star F+(d\alpha)\wedge F=J\,,\quad dF=0\,.
\end{equation} 
It is as if the current $J$ picked up an additional piece depending on
the gradient of the axion field.  The real part of Kiehn's {\it chiral
  vacuum\/} \cite{Kiehn2002} is a subcase, for $\alpha={\rm const}$,
of axion electrodynamics.

Decomposing the inhomogeneous equation in time and space, we have ($c$
is the speed of light and $^{\underline{*}}$ the 3-dimensional Hodge
star)
\begin{eqnarray}\label{Minhom1}
  \frac{\lambda_0}{c}\, \underline{d}\,^{\underline{*}}E &=&
  \rho-(\underline{d}\alpha)\wedge B\,,\\ 
  \lambda_0\hspace{1pt}c\,\underline{d}\,^{\underline{*}}B -
  \frac{\lambda_0}{c}\,\partial_t{^{\underline{*}}E}& =&
  j+(\underline{d}\alpha)\wedge
  E+(\partial_t{\alpha})\,B\,.\label{Minhom2}
\end{eqnarray}
Possible experimental verifications were suggested by Sikivie and
others, see \cite{Sikivie1983,Sikivie1999,Cern}.

\subsubsection*{Pure axion field and the gyrator}

As a degenerate special case, we can also consider the {pure
  (``stand-alone'') axion\/} field with $\,^{(1)}\kappa_{ij}{}^{kl} =
\,^{(2)}\kappa_{ij}{}^{kl}=0$. Then,
\begin{equation}\label{standalone1}
  H=\alpha\,F\qquad{\rm or}\qquad {\cal
    H}=-\alpha\,E\,,\quad {\cal D}=\alpha\,B\,,
\end{equation}
and the Maxwell equations read
\begin{equation}\label{standalone2}
  (d\alpha)\wedge F=J\qquad{\rm and}\qquad dF=0\,.
\end{equation}
This is a special case of axion electrodynamics, namely (\ref{MaxAx})
with $\lambda_0=0$. Here we have a case in which the constitutive
matrix $\frak B$ in (\ref{const1}) vanishes. Consequently, $\det
{\frak B}=0$ and the $E{\cal H}$-version (\ref{neu6}),(\ref{neu7}) of
the constitutive law loses its meaning.

Seemingly the first person to discuss (and to reject) a constant pure
axion field was Schr\"odinger \cite{Schroedinger}, p.25, penultimate
paragraph, and, as a non-constant field, Dicke \cite{Dicke64}. The
framework (\ref{standalone1}),(\ref{standalone2}) corresponds to
Tellegen's {\it gyrator\/} \cite{Tellegen1948,Tellegen1956/7} and to
Lindell \& Sihvola's {\it perfect electromagnetic conductor\/} (PEMC)
\cite{LindSihv2004a,LindSihv2004b}.

Lindell \& Sihvola define their PEMC by using directly the relation
$H=\alpha F$.  Tellegen \cite{Tellegen1948,Tellegen1956/7} considered
a linear network of two terminal pairs (four-pole or two-port):
\begin{eqnarray}\label{gy1}
  v_1 & =& a_{11}\, i_1 + a_{12}\, i_2\, ,\\ 
  v_2 & =& a_{21}\, i_1 + a_{22}\, i_2\, .
\end{eqnarray}
The v's represent voltages, the i's currents, and the (frequency
dependent) a's resistances.  Since the field strengths are related to
the voltage ($E\stackrel{{\rm SI}}{\sim} V/m$, $B\stackrel{{\rm
    SI}}{\sim} Vs/m^2$) and the excitations to the current (${\cal
  H}\stackrel{{\rm SI}}{\sim} A/m$, ${\cal D}\stackrel{{\rm SI}}{\sim}
As/m^2$), {\it our\/} translation into the linear medium picture
reads:
\begin{eqnarray}\label{gy2}
  E & =& a_{11}\, {\cal D} + a_{12}\,{\cal H}\, ,\\ 
  B & =& a_{21}\,  {\cal D} + a_{22}\,{\cal H}\,  .
\end{eqnarray}
Tellegen had a different translation, since he used the $E{\cal
  H}$-formalism of (\ref{neu6}),(\ref{neu7}).  Tellegen
\cite{Tellegen1948}, eq.(10), defined his gyrator by
\begin{eqnarray}\label{gy3}
  v_1 & =& -\sigma\, i_2\,,\\ v_2 & =&\hspace{9pt} \sigma\, i_1 \,.
\end{eqnarray}
The gyrator ``rotates'' currents into voltages and vice versa. In our
medium picture we have
\begin{eqnarray}\label{gy3'}
  E & =& -\sigma\,{\cal H}\,,\\ B & =&\hspace{9pt} \sigma\,{\cal D}
  \,,\label{gy3''}
\end{eqnarray}
i.e., the excitations are rotated into the field strengths. As a
comparison with (\ref{standalone1}) shows, this corresponds to the
axion field. Lindell \& Sihvola \cite{LindSihv2004a} have shown that
by using the $E{\cal H}$-formalism and by performing a suitable
limiting transition (since $\det \frak B=0$), one can also arrive at
(\ref{gy3'}),(\ref{gy3''}).  Consequently, it is established that the
medium analogue of Tellegen's gyrator is the axion field --- and both
are equivalent to the PEMC of Lindell \& Sihvola.

The gyrator has been first realized for microwaves by the use of
premagnetized ferrites. A corresponding figure with detailed
explanations can be found in the review article of Hogan \cite{Hogan},
Fig.12.

\section{The constitutive tensor $\chi^{ijkl}$ and its properties}

We can raise the two lower indices of the constitutive tensor
$\kappa_{ij}{}^{kl}$ with the help of the contravariant Levi--Civita
symbol ${\epsilon}^{ijmn}$. Thus, we can introduce the constitutive
tensor {\em density} of spacetime,
\begin{equation}\label{raise}
  \chi^{ijkl}:= \frac{1}{2}\,{\epsilon}^{ijmn}\, \kappa_{mn}{}^{kl}
  \,,
\end{equation}
which is equivalent to $\kappa_{ij}{}^{kl}$ but more widely in use. The
notation $\chi$ was chosen in order to conform with Post's convention
\cite{Post}. The constitutive law can now be written in the
conventional form as
\begin{equation}\label{constPost}
  \check{H}{}^{ij}=\frac{1}{2}\,\chi^{ijkl}\,F_{kl}\,,
\end{equation}
see Post \cite{Post}, eq.(6.12). The constitutive tensor with all 36
components has also been discussed in an equivalent way by Lindell in
his book \cite{Ismo}, see also
\cite{Schultz,KiehnFresnel,SihvolaLindell1995,RaabSihvola1997,LindellWallen2004a,LindellWallen2004b,LindSihv2004a,LindSihv2004b}.

We can decompose $\chi^{ijkl}$ irreducibly under the linear group,
too. With the expressions
\begin{eqnarray}
  {}^{(1)}\chi^{ijkl} &=& {\frac 1 2}\,\epsilon^{ijmn}\,\,{}^{(1)}
  \kappa_{mn}{}^{kl},\label{chi1}\\ {}^{(2)}\chi^{ijkl} &=& {\frac 1
    2}\,\epsilon^{ijmn}\,\,{}^{(2)} \kappa_{mn}{}^{kl} =
  -\,\epsilon^{ijm[k}\!\not\!\kappa_m{}^{l]},\label{chi2}\\ 
  {}^{(3)}\chi^{ijkl} &=& {\frac 1 2}\,\epsilon^{ijmn}\,\,{}^{(3)}
  \kappa_{mn}{}^{kl} = {\frac 1 {12}}\,\epsilon^{ijkl}\,\kappa\,,
  \label{chi3}
\end{eqnarray}
we find\footnote{We want to underline that the irreducible
  decomposition of $\chi^{ijkl}$ is {\it algebraically\/} similar the
  decomposition of a curvature tensor in a Riemann-Cartan space
  (Riemannian space with a metric compatible connection). Of course,
  only in a Riemann-Cartan space the curvature tensor has 36
  independent components and can carry a totally antisymmetric piece
  $R^{[ijkl]}$. This piece, by the first Bianchi identity, is directly
  related to Cartan's torsion. But let us stress, these are just
  algebraic analogies, not more and not less.  Premetric
  electrodynamics, as represented by eqs.(1) and (2), does neither
  couple to the metric nor to the connection of spacetime.  Thus also
  torsion is not involved. Moreover, the spacetime relation eventually
  defines a metric, but that's all.}
\begin{equation}
  \underbrace{ \chi^{ijkl}}_{36} =
  \underbrace{{}^{(1)}\chi^{ijkl}}_{principal\> 20} +\underbrace{
    {}^{(2)}\chi^{ijkl}}_{skewon\> 15} + \underbrace{
    {}^{(3)}\chi^{ijkl}}_{axion\> 1}\label{chi-dec}\,.
\end{equation}
The irreducible pieces carry the additional symmetries
\begin{equation}\label{symm}
 {}^{(1)}\chi^{ijkl}= {}^{(1)}\chi^{klij}\,,\quad 
 {}^{(2)}\chi^{ijkl}=- {}^{(2)}\chi^{klij}\,,\quad
 {}^{(3)}\chi^{ijkl}= {}^{(3)}\chi^{[ijkl]}\,.
\end{equation}

Clearly, the {\it principal\/} part with its 20 independent components
is the one discussed by Post \cite{Post}. The {\it skewon\/} part with
its 15 components vanishes if one assumes the existence of a
Lagrangian 4-form from which the spacetime relation can be derived in
full, see below.  However, this assumption is not a necessary, as we
have shown \cite{Birkbook}. Finally, the {\it axion\/} piece with only
1 independent component is totally antisymmetric:
\begin{equation}
 {}^{(3)}\chi^{ijkl}=\a\,\epsilon^{ijkl}\,.
\end{equation}
In Post's book \cite{Post}, eq.(6.18), this piece is forbidden;
therefore, Lakhtakia and Weiglhofer \cite{LakhWeig1994} called
it the Post constraint:
\begin{equation}
  {}^{(3)}\chi^{ijkl}= 0\qquad{\rm or}\qquad \alpha=0\qquad{\rm
    (Post\>constraint)}\,.
\end{equation}
Post even carries on and requires $\partial_i\alpha=0$, see his
eq.(6.19).  The reasons he gave for the latter constraint are
incomprehensible to us. Therefore, we will not dwell on it any
further.

We want now to turn to criteria that will help us to decide which
pieces of the constitutive tensor are reasonable from a physical point
of view. Our kinematical analysis so far gave no convincing argument
why any of its 36 components should not exist in nature. In the
sequel, we will distinguish between case (i), the vacuum, and case (ii),
material media.

In electrodynamics, after formulating the Maxwell equations, one has
to specify the energy-momentum density of the electromagnetic field.
Also this can be done in a metric-free environment, see
\cite{Birkbook}. The {\it energy-momentum\/} 3-form in (holonomic)
coordinates reads
\begin{equation}\label{energy}
  \Sigma_i=\frac{1}{2}\left(F\wedge e_i\rfloor H -
    H\wedge e_i\rfloor F\right)\,,
\end{equation}
with $e_i=\partial_i$ as the 4-dimensional vector basis of the tangent
space. Let us denote the basis of the 4-dimensional 3-forms by
$\hat{\epsilon_k}=\epsilon_{klmn}\,dx^l\wedge dx^m\wedge dx^n/6$. If we
then decompose the energy-momentum 3-form with respect to
$\hat{\epsilon_k}$, that is, $\Sigma_i= {\cal
  T}_i{}^{k}\,\hat{\epsilon_k}$, we find the conventional Minkowski
energy-momentum tensor
\begin{equation}\label{ff}
  {\cal T}_i{}^{k}=\frac{1}{4}\delta_i^k
  F_{lm}\check{{H}}^{lm}-F_{il}\check{{H}}^{kl}\,.
\end{equation}
Of course, (\ref{energy}) and (\ref{ff}) are equivalent
representations of the energy-momen\-tum distribution.

The representation (\ref{energy}) is most convenient for showing that
the axion piece, even as field, doesn't carry electromagnetic
energy-momentum. The constitutive law $H=\alpha(x) F$ yields $F\wedge
e_i\rfloor H=F\wedge e_i\rfloor(\alpha F)=\alpha F\wedge e_i\rfloor
F$.  Together with the second piece in (\ref{energy}), this vanishes.
For the principal piece it can be shown that it leads to positive
energy, see \cite{ItinHehl}, also the skewon piece has non-vanishing
contributions in general, see Table 1. We are talking here about the
{\it electromagnetic\/} energy. If the axion field contributed an own
kinetic term \`a la $\sim g^{ik}(\partial_i \alpha)(\partial_k
\alpha)/2$ to the Lagrangian, then it would carry own energy.

\renewcommand{\arraystretch}{1.5}
\begin{table}[ht]\label{contributes}
\centering
\caption{Contribution of each irreducible part $^{(r)}\chi$ of the 
  constitutive tensor density to the electromagnetic energy-momentum,
  to the Lagrangian, and to the {\bf T}amm-{\bf R}ubilar tensor
  density, see \cite{Birkbook}. Our results are consistent with the
  findings of Kiehn \cite{Kiehn2002,Kiehn2003}.}\bigskip

\begin{tabular}{|c||c|c|c|}\hline
  \multicolumn{1}{|c||} {Irreducible} &\multicolumn{3}{|c|}
  {Contributes to}\\ \cline{2-4} part & energy-momentum\ & Lagrangian
  & TR tensor density\\ \hline principal $^{(1)}\chi\quad\;$ & yes &
  yes & yes \\ \hline skewon $^{(2)}\chi$ or $\!\not\!S$ & yes & no &
  yes \\ \hline axion $^{(3)}\chi$ or $\alpha$ & no & yes & no \\ 
  \hline
\end{tabular}\medskip

\end{table}
\renewcommand{\arraystretch}{1}

This brings us to the Lagrangian of the electromagnetic field. The
Maxwell equations (\ref{inhMax}),(\ref{homMax}) can be derived from
electric charge and magnetic flux conservation and the energy-momentum
current (\ref{energy}) via the Lorentz force density. A Lagrangian is
not needed in this procedure. The same it true for the local and
linear constitutive law (in the sense of irreversible thermodynamics).
However, one can find an electromagnetic {\it Lagrangian\/} 4-form
\begin{equation}\label{Lagrangian}
V=-\frac{1}{2}\,H\wedge F\,.
\end{equation}
By substituting $H=\kappa(F)$ into the Lagrangian, the piece
$^{(2)}\kappa(F)$ drops out because of the ``anti-Onsager'' symmetry
(\ref{symm})$_2$. This shows that a possible skewon piece is related
to {\it dissipative\/} effects. For material media this is certainly a
possibility. For empty spacetime it may be a legitimate hypothesis.

We will now turn to the propagation of the electromagnetic field (of
``light''). In the geometrical optics limit, we find for the wave
covector $q$ the Fresnel equation \cite{Fukui,Birkbook}
\begin{equation} \label{Fresnel}   
 {{\cal G}^{ijkl}(\chi)\,q_i q_j q_k q_l = 0 \,,}  
\end{equation} 
with the (metric-free) Tamm-Rubilar tensor density 
\begin{equation}\label{G4}   
  {{\cal G}^{ijkl}(\chi):=\frac{1}{4!}\,\hat{\epsilon}_{mnpq}\, 
  \hat{\epsilon}_{rstu}\, {\chi}^{mnr(i}\, {\chi}^{j|ps|k}\, 
  {\chi}^{l)qtu }\,.} 
\end{equation} 
By simple algebra it is possible to prove that $\chi^{[ijkl]}$ drops
out from the Tamm-Rubilar tensor: ${\cal
  G}^{ijkl}(^{(1)}\chi+\,^{(2)}\chi+\,^{(3)}\chi)={\cal
  G}^{ijkl}(^{(1)}\chi+\,\chi^{(2)})$. Furthermore, ${\cal
  G}^{ijkl}(^{(3)}\chi)=0$. In other words, the axion part
$^{(3)}\chi^{ijkl}=\epsilon^{ijkl}\alpha(x)$, in the
geometrical optics limit, does {\it not\/} influence the propagation
of light. For the skewon piece we find ${\cal
  G}^{ijkl}(^{(2)}\chi)=0$. But even more far-reaching, by
straightforward algebra, see \cite{Birkbook}, it can be shown that
${\cal G}^{ijkl}(^{(2)}\chi+\,^{(3)}\chi)=0$.  Consequently, for
$^{(1)}\chi=0$, the Fresnel equation collapses and there is no orderly
wave propagation in the geometrical optics limit.  Therefore, the
existence of conventional wave propagation requires $^{(1)}\chi\ne 0$,
that is, the principal part of the constitutive tensor must be
non-vanishing.

Although from Table~1 one may have an impression that the axion field
is to a great extent irrelevant in experiments, it {\it does} affect
physical observations. In particular, the important phenomenon of
electromagnetic waves is certainly not exhausted by geometrical
optics. Specifically, in the framework of axion electrodynamics
(\ref{MaxAx}) with a constant principal part $^{(1)}\chi^{ijkl}$, Itin
\cite{Yakov2004} (see also Carroll et al.\ \cite{carroll}) has shown
that the axion field does influence electromagnetic waves if one looks
into exact solutions. Assuming the usual plane wave ansatz for the
electromagnetic field strength $F = f\,e^{i\varphi}$, with the
constant amplitude 2-form $f$ and the phase function $\varphi(x^i)$,
we find a generalization of the Fresnel equation (\ref{Fresnel}) in
the form of \cite{Yakov2004}
\begin{equation} \label{FresnelAxi} 
{\cal G}^{ijkl}(\chi)\,q_i q_j q_k
q_l - \chi^{ijkl}(\partial_i\alpha)(\partial_l\alpha)\,q_j q_k = 0\,.
\end{equation}
Here $q_i = \partial_i\varphi$ is the wave covector, as usual. This
extended Fresnel equation is no longer an homogeneous algebraic
equation, and the axion field contributes explicitly to the second
term.  This necessarily leads, in the presence of the axion field, to
the birefringence of the waves.

As another manifestation of the axion field, let us consider the
Lorentz force density $f_i$ and the balance equation related to it
(see \cite{Birkbook}, Sec.~B5.1):
\begin{equation}
f_i = d\Sigma_i + X_i\,,\quad {\rm with}\quad X_i := -\,{\frac 12}\left(
F\wedge {\cal L}_iH -  H\wedge {\cal L}_iF\right)\,.
\end{equation}
This is the force density that an electromagnetic field exerts on
charges and currents.  The axion field indeed drops out from the first
term, which contains the electromagnetic energy-momentum current
(\ref{energy}). However, the second term is nontrivial.  Using the
elementary properties of the Lie derivative ${\cal L}_i$, we
straightforwardly find for the constitutive law $H=\lambda_0\,^\star
F+\alpha F$,
\begin{equation}
X_i = -\,{\frac 12}(\partial_i\alpha)\,F\wedge F\,.
\end{equation}
This additional force density arises from the axion
field.\footnote{There are proposals to combine the latter force term
  with the usual term $d\Sigma_i$, thus arriving at a modified
  energy-momentum current \cite{carroll,Yakov2004}.  However, this
  only works under the additional assumption of the constancy of the
  gradient of the axion field, $d(\partial_i\alpha) =0$, and results
  in a non gauge invariant expression for the energy-momentum.}

\section{Constitutive tensor split in space and time}

We will now come back to the constitutive relations
(\ref{const1}),(\ref{const2}). We will again work with a 3--vector
formalism. Again, we formulate it in a {\it metric independent\/} way.
We decompose the electromagnetic field with respect to the
3--dimensional coframe $\vartheta^a$ and the 2--form basis
$\hat{\epsilon}_b={\hat{\epsilon}}_{bcd}\,\vartheta^c\wedge
\vartheta^d/2$:
\begin{eqnarray}
  {\cal H}&=& {\cal H}_a\, \vartheta^a\,, \quad E=E_a\, \vartheta^a\,,
  \\ {\cal D}&=& { {\cal D}}^b\,\, \hat{\epsilon}_b\,,\quad\>
  B={B}^b\, \hat{\epsilon}_b\,.
\end{eqnarray}
Then the constitutive relations (\ref{const1}),(\ref{const2}) can be
put in the form of a matrix equation \cite{Birkbook},
\begin{equation}
  \left(\begin{array}{c} {\cal H}_a \\ {\cal D}^a\end{array}\right) =
  \left(\begin{array}{cc} {{\frak C}}^{b}{}_a & {{\frak B}}_{ba} \\ 
      {{\frak A}}^{ba}& {{\frak D}}_{b}{}^a \end{array}\right)
  \left(\begin{array}{c} -E_b\\ {B}^b\end{array}\right)=:
  \left(\begin{array}{c}\hspace{-8pt}\null\\ \hspace{-8pt}\null
    \end{array}\right)\hspace{-15pt}\kappa \;\,
  \left(\begin{array}{c} -E_b\\ {B}^b\end{array}\right)\,.\label{CR'}
\end{equation}
Moreover, by using the Levi-Civita symbol, see (\ref{raise}), we find
for the $\chi$ tensor density
\begin{equation}\label{kappachi}
 \left(\begin{array}{c}\hspace{-8pt}\null\\ \hspace{-8pt}\null
    \end{array}\right)\hspace{-15pt}\chi \;\,= \left( \begin{array}{cc}
      {\cal B}_{ab}& {\cal D}_a{}^b \\ {\cal C}^a{}_b & {\cal A}^{ab}
    \end{array}\right)\,.
\end{equation}
If we compare (\ref{kappachi}) with (\ref{constit})$_1$, then, by
straightforward algebra, the constitutive $3\times 3$ matrices ${\cal
  A,B,C,D}$ can be related to the 4-dimensional constitutive tensor
density (\ref{raise}) by
\begin{eqnarray}\label{A-matrix0}
  {\frak A}^{ba}&=& \chi^{0a0b}\,,\quad
{\frak B}_{ba}=\frac{1}{4}\,\hat\epsilon_{acd}\,
\hat\epsilon_{be\!f} \,\chi^{cdef}\,,\\
\label{C-matrix0}
{\frak C}^a{}_b& =&\frac{1}{2}\,\hat\epsilon_{bcd}\,\chi^{cd0a}\,,\quad
{\frak D}_a{}^b=\frac{1}{2}\,\hat\epsilon_{acd}
\,\chi^{0bcd}\,.
\end{eqnarray}

Let us now employ the irreducible decomposition. For the principal
part of $\kappa$ or $\chi$, respectively, we take the notation of Post
\cite{Post}, namely $\varepsilon^{ab}=\varepsilon^{ba}$ (6 independent
components), $\mu^{-1}_{ab}=\mu^{-1}_{ba}$ (6 components), and
$\gamma^a{}_b$, with $\gamma^c{}_c=0$ (8 components). The skewon part
we parametrize according to
\begin{equation}\label{prrameterS}
  \!\not\!S_i{}^j= \left(\begin{array}{cc}-s_c{}^c & m^a \\ n_b &
      s_b{}^a \end{array}\right)\,,
\end{equation}
with, in 3 dimensions, the tensor skewon field $s_a{}^b$ (9
independent components), the vector skewon $m^a$ (3 components), and
the covector skewon $n_a$ (3 components). If we insert all of this in
$\kappa$, we eventually find
\begin{equation}
 \left(\begin{array}{c}\hspace{-8pt}\null\\ \hspace{-8pt}\null
    \end{array}\right)\hspace{-15pt}\kappa \;\,
  = \underbrace{\left(\begin{array}{cc}
        \gamma^b{}_a & \mu_{ab}^{-1} \\ -\varepsilon^{ab} &
        \gamma^a{}_b \end{array}\right)}_{ principal\>{\rm part\> 20\>
      comp.}} + \underbrace{ \left(\begin{array}{cc} - s_a{}^b
        +\delta_a^b s_c{}^c &\>\; - \hat{\epsilon}_{abc}m^c \\ 
        \epsilon^{abc}n_c & s_b{}^a - \delta_b^a s_c{}^c
  \end{array}\right)}_{ skewon\> {\rm part\> 15\> comp.}}
   + \underbrace{\alpha\left(\begin{array}{cc}\delta_a^b&0\\0&\delta_b^a
\end{array}\right)}_{ axion\>{\rm part\> 1 \>comp.}}.\label{CR''''}
\end{equation}
According to (\ref{CR'}), we can evaluate the matrix elements:
\begin{eqnarray}\label{explicit'}
  {\cal H}_a\!&=\!&\left( \mu_{ab}^{-1} - \hat{\epsilon}_{abc}m^c
  \right) {B}^b +\left(- \gamma^b{}_a + s_a{}^b - \delta_a^b
    s_c{}^c\right)E_b - \alpha\,E_a\,,\\ {\cal D}^a\!&=\!&\left(
    \varepsilon^{ab}\hspace{4pt} - \, \epsilon^{abc}\,n_c
  \right)E_b\,+\left(\hspace{9pt} \gamma^a{}_b + s_b{}^a - \delta_b^a
    s_c{}^c\right) {B}^b + \alpha\,B^a \,.\label{explicit''}
\end{eqnarray}
Recall that $\varepsilon^{ab}=\varepsilon^{ba}$,
$\mu^{-1}_{ab}=\mu^{-1}_{ba}$, and $\gamma^c{}_c=0$. Incidentally,
$\alpha$ is a 4-dimensional (axial) scalar, whereas $s_c{}^c$ is only
  a 3-dimensional scalar. The cross-term $\gamma^a{}_b$ is related to
  the Fresnel-Fizeau effects.  The skewon contributions $m^c,n_c$ are
  responsible for electric and magnetic Faraday effects, respectively,
  whereas the skewon terms $s_a{}^b$ describe optical activity.

\section{Constitutive tensor and physics}\label{physics}

Up to here, we haven't said much about which piece of the constitutive
tensor may be realized in nature. We just made a ``kinematical''
investigation of the most general local and linear constitutive tensor
with its 36 independent components. From an analysis of the Tamm-Rubilar
tensor we found that $ {}^{(1)}\chi^{ijkl}\ne 0 $ is necessary for a
non-degenerate propagation of electromagnetic waves in the geometrical
optics limit.  Only then the Tamm-Rubilar tensor doesn't vanish and
the Fresnel equation exists. By the same method we find that, in the
same limit, $ {}^{(3)}\chi^{ijkl}\ne 0$ doesn't disturb the wave
propagation at all.

The {\it skewon\/} part $ {}^{(2)}\chi^{ijkl}$, in its full
generality, was discussed, amongst others, by Sihvola \& Lindell
\cite{SihvolaLindell1995,Ismo} and by us \cite{Birkbook}. We studied
already the possible effects of the skewon field on light propagation
\cite{skewon}. But we should mention Nieves and Pal \cite{NP94} who
introduced, besides the vacuum impedance $1/\lambda_0$ and the
velocity of light $c$, a third constant for the vacuum by means of a
specially chosen skewon piece.  Nieves and Pal assume (in our
notation)
\begin{equation}\label{eq30}
  s_a{}^b= \frac{s}{2}\,\delta_a^b\,,\quad m^a=0\,,\quad n_a=0\quad{\rm
    (Nieves\;\&\;Pal)}\,.
\end{equation}
This corresponds to a spatially isotropic skewon field. By
(\ref{eq30}), a constraint is given that restricts the class of
allowed reference frames. Lakhtakia\footnote{Private communication.}
calls it {\it ``isotropic chirality which is Lorentz reciprocal''}.
The Post constraint doesn't forbid it since the Nieves \& Pal piece
relates to the skewon part, whereas the violation of the Post
constraint to the axion part.  Lakhtakia points out that $s\ne 0$ is
valid for certain {\it materials,\/} but not for vacuum.  Furthermore,
others (see \cite{Birkbook}) put forward the hypothesis that there can
be a fourth electromagnetic constant that makes the medium {\it
  ``nonreciprocal (but still isotropic)''}.  It is, of course, the
axion field. With (\ref{eq30}), the constitutive law
(\ref{explicit'}),(\ref{explicit''}) reads
\begin{align}\label{resultx}
  {\cal H}_a &=- \left(s+ \alpha\right)E_a+\,^{(1)}{\cal
    H}_a(E,{B})\,, \\ {{\cal D}}^a &=\;^{(1)} {{\cal D}}^a(E,{B})
  -\left(s- \alpha\right){B}^a\,,\label{resulty}
\end{align}
where we used the notation $^{(1)}{\cal H}:=\,^{(1)}\kappa(F)$ and
$^{(1)}{\cal D}:=\,^{(1)}\kappa(F)$. We rewrite
(\ref{resultx}),(\ref{resulty}) for the case of a Maxwell-Lorentzian
principal piece in exterior calculus as
\begin{align} 
  {\cal H} &=\mu^{-1}\,^{\underline{*}}B- \left(s+ \alpha\right)E \,,
  \\ {{\cal D}} &=\;\;\varepsilon\,\,^{\underline{*}}E\; -\left(s-
    \alpha\right){B} \,.
\end{align}
Note how the isotropic skewon field and the axion field act with
different signs on $E_a$ and ${B}^a$. Thereby they can be
distinguished phenomenologically. We stress that the axion field
$\alpha$ is a premetric, 4-dimensional scalar with twist and as such
of fundamental potential importance. In contrast, the skewon field $s$
is only a 3-dimensional scalar valid in a constrained class of
reference frames.---

What is then the {\it the state of the Post constraint?\/} Of course,
the somewhat abstract concepts of a pure axion field, the Tellegen
gyrator, and the PEMC of Lindell \& Sihvola all violate the Post
constraint, that is, $\alpha\ne 0$, whereas the other pieces of the
constitutive tensor vanish at the same time.

One argument in favor of the Post constraint is the following (see
\cite{Weig1994}): With $^{(1)}H:=\,^{(1)}\kappa(F)$ and
$^{(2)}H:=\,^{(2)}\kappa(F)$, the inhomogeneous Maxwell equation for
the linear constitutive law reads
\begin{equation}\label{inhalpha}
dH=d\left(^{(1)}H+{}^{(2)}H \right)+(d\alpha)\wedge F=J\,,
\end{equation}
where we used already the homogeneous Maxwell equation $dF=0$, see
also (\ref{MaxAx}). If the axion piece of the constitutive tensor is
{\it constant,\/} i.e., $d\alpha =0$, then $\alpha$ cannot couple to
the Maxwell equations.  However, this is only true for an infinitely
extended material. As soon as the material has finite size, the axion
piece $\alpha$ jumps at the surfaces of the material --- and this jump
of $\alpha$ couples to the Maxwell equations. In other words, the
axion field can be seen along those jump surfaces by means of
electromagnetic waves beyond the geometrical optics limit. In order to
make this explicit, let us assume that a two-dimensional surface $S$
divides the space into two parts. Let now both half-spaces be
filled with homogeneous material media both of which are characterized
by constant but unequal values of the axion field: $\alpha_1$,
say, in the first material and $\alpha_2$ in the second one. Then, if
we take as an example the pure gyrator case with the constitutive law
(\ref{standalone1}), we find from the jump conditions on the boundary
(see \cite{Birkbook}, Sec.~B.4.3),
\begin{equation}\label{jump}
[\alpha]\,n_a B^a\,\vline\,{\hbox{\raisebox{-1.5ex}{\scriptsize{S}}}} =
\widetilde{\rho}\,,\qquad [\alpha]\,\epsilon^{abc} n_b E_c\,\vline
\,{\hbox{\raisebox{-1.5ex}{\scriptsize{S}}}} = -\,\widetilde{j}^a\,.
\end{equation}
Here $n_a$ are the components of the covector normal to the surface
$S$, and $[\alpha] := \alpha_2 - \alpha_1$ is the jump of the axion
field on the boundary. Thus, even though the constant axion drops out
from the Maxwell equations inside both homogeneous regions, it pops up
in the form of the surface charge density $\widetilde{\rho}$ and the
surface current density $\widetilde{j}^a$ induced on the boundary $S$.
Both are proportional to the jump of the axion $[\alpha]$. In the more
general case of axion electrodynamics, see (\ref{MaxAx}), the
left-hand sides in (\ref{jump}) will include the jump of the normal
components of the electric excitation ${\cal D}^a$ and the jump of the
tangential components of the magnetic excitation ${\cal H}_a$,
respectively. Hence our argument can be generalized straightforwardly.

This enables us to measure the value of the axion field of a
homogeneous material medium.  In this sense, the axion field of a
material medium, our case (ii), always couples to the inhomogeneous
Maxwell equation, and the argument mentioned above no longer applies.
In the case of spacetime, also a curved one, the argument is correct
if really everywhere $d\alpha =0$.  As soon as the axion field becomes
space and/or time dependent, also in case (i) the quoted argument
loses its meaning.

For empty spacetime, our case (i), the Post constraint, according to
experimental evidence, is fulfilled. However, the relaxation of the
Post constraint leads to the concept of the axion field. According to
our Table 1, this field is very elusive since it neither shows up in
the electromagnetic energy-momentum tensor nor in the
Tamm-Rubilar-tensor, i.e., it doesn't influence light propagation in
the limit of geometrical optics. Hence it is hard to detect it.  A
whole industry is at work to find the axion field, see the new data on
elementary particles \cite{Cern}, p.\ 389. So far without success.
There is no theoretical reason known to us that would forbid the axion
field.  Quite the opposite. From a theoretical point of view, it
appears as a quite natural extension of Maxwell-Lorentz
electrodynamics. But, as mentioned, the axion hasn't been found so
far.

For material media, our case (ii), the situation is different.
Gyrators have been engineered, see \cite{Hogan}. Consequently, there
is little doubt that materials can be constructed that violate the
Post constraint. The Tellegen model with parallel or antiparallel
electric and magnetic dipoles looks reasonable, see Tretyakov et al.\ 
\cite{Tretyakov1998}. In the meantime, Tretyakov et al.\ 
\cite{Tretyakov2003} built an artificial {\it Tellegen particle\/} and
verified its nonreciprocal magnetoelectric behavior.  ``Natural''
material media exist that violate the Post constraint, namely
Cr$_2$O$_3$ and in Fe$_2$Te$\,$O$_6$ in static magnetic fields, see de
Lange \& Raab \cite{LangeRaab}.  Accordingly, model calculations
\cite{Ponti2002,RaabSihvola1997} and experiments \cite{LangeRaab} (see
also \cite{Raab}) show that the Post constraint is valid for many
materials; however, it is definitely {\it violated\/} in some rare
cases. Thus, the Post constraint as a general dogma should be buried
with all due honors.

\section*{Acknowledgment}

We would like to thank R.M.~Kiehn, Ismo Lindell, and E.~Jan Post for
numerous fruitful discussions via email. We also appreciate having
received from them pre- and reprints of their and others work.  We are
grateful to Frank Gronwald, Yakov Itin, and Roger Raab for helpful
remarks.  Financial support from the DFG (HE-528/20-1) is gratefully
acknowledged.

\begin{footnotesize}

\centerline{=========}
\end{footnotesize}

\end{document}